\documentstyle[preprint,aps]{revtex}
\begin{document}
\draft

%\preprint{FTUV/96-56; IFIC/96-64}
\title{Meson Exchange Currents in Kaon Scattering 
on the Lightest Nuclei}
\author{ S. S. Kamalov\cite{Sabit}, J. A. Oller, E. Oset and 
M. J. Vicente-Vacas}
\address{ Departamento de F\'{\i}sica Te\'orica and IFIC, Centro Mixto 
Universidad de Valencia - CSIC, 46100 Burjassot (Valencia) Spain.}

\date{\today}
\maketitle

\begin{abstract}
The $K^+$ scattering on the lightest nuclei, $d$, $^3He$ and 
$^4He$ is studied in the framework of multiple scattering 
theory.  Effects from  MEC tied to the $K^+N\rightarrow KN\pi$ 
reaction are evaluated. We found that at momentum transfers 
$Q^2<0.5$ (GeV/c)$^2$ contributions from  MEC are much smaller 
than kaon rescattering corrections.  This  makes the 
conventional multiple scattering picture a reliable tool to 
study these reactions in this kinematical domain and to extract 
the $K^+n$ scattering amplitude from the $K^+d$ data. At larger 
transferred momentum MEC can become more relevant.  
\end{abstract}


\section{INTRODUCTION}

The $K^+$-nucleon interaction is one of the weakest among other 
strongly interacting systems. Therefore, it is expected that for 
the description of the kaon-nuclear interaction a simple  
Impulse Approximation (IA) should be more reliable.  However, 
recent theoretical 
\cite{SKG,Weise,Chen,Ernst,Koltun,Oset1,Meiss}  studies in this 
framework show systematic discrepancies  with the experimental 
data, in particular, for the ratio of the  $K^+$-nuclear to 
$K^+$-deuteron total cross sections. This puzzling situation 
still persists in the description of $K^+-^{12}C$ and 
$K^+-^{40}Ca$ scattering and most of the theoretical and 
experimental studies are concentrated on these targets. The 
reader can  find a review of this problem in Ref.\cite{Ernst}.  
A more recent experimental and theoretical study of $K^+$ 
scattering on $C$ (natural) and $^6Li$ shows agreement between 
theory and experiment within errors for the elastic cross 
sections, but discrepancies for the ratio of the total cross 
sections where systematic errors can be eliminated \cite{Mich}.  
On the other hand, the systematic study of the $K^+$ interaction 
with the lightest nuclei (deuteron, $^3He$, and $^4He$) could 
provide us with new knowledge about the dynamics of the 
kaon-nuclear interaction, which could be applied then to the 
case of heavier nuclei. Examples of practical realization of 
such scheme were done in the study of pion-nuclear 
\cite{Landau75,Mach76,LT78,KTB93,KTB95}  and 
kaon-nuclear~\cite{Paez} scattering  within multiple scattering 
theory. In the present paper we will extend it by including 
Meson Exchange Current (MEC) corrections.

From the study of electromagnetic interactions it is well known 
that at high momentum transfers MEC play an important role and it 
is important to find out their relevance in $K^+$ nucleus 
scattering.  In fact, the determination of the $K^+n$  amplitude 
from the study of the of $K^+$-deuteron scattering is tied to 
the hypothetical strength of the MEC. This point is stressed in 
Ref.\cite{Paez} where the uncertainty in the $K^+$ nucleus 
cross section from likely effects of MEC is repeatedly mentioned 
and kept in mind as a possible source of corrections needed to 
explain systematic discrepancies of the calculations with the 
data.

The first attempts to calculate MEC effects in $K^+$ nuclear 
scattering were done for nuclear matter~\cite{Koltun,Oset1}. 
Remaining uncertainties in Refs.\cite{Koltun,Oset1} tied to the 
off-shell extrapolation of the $K \pi$ amplitude were settled in 
Ref.\cite{Meiss} using the Chiral Lagrangians involving the octet 
of pseudoscalar mesons and nucleon and delta  in the baryon 
sector. The calculations in Ref.\cite{Meiss} were also done in 
spin-isospin saturated nuclear matter where some cancellations 
occur due to the $T=0$ and $J=0$ character of the medium.  In 
the present paper we shall evaluate for the first time the MEC 
using nuclear wave functions with detailed  spin-isospin 
structure.  As a starting point to construct the MEC operator 
for kaon-nuclear scattering we shall use the amplitude for the 
$K^+N\rightarrow KN\pi$ reaction obtained recently in 
Ref.\cite{Oset2} in the framework of standard Chiral 
Perturbation Theory. 

Finally let us make one comment about $K^+d$ scattering.  As we 
mentioned above, this reaction is the main source of information 
about $K^+$ interaction with the neutron. From the study of 
pion-deuteron scattering we know that, for this purpose, the 
three-body Faddeev approach would be the best theoretical tool 
which offers the possibility to treat the $NN$ interaction and 
scattering processes within a unified and consistent framework. 
Fortunately, in the case of $K^+NN$ system, due to the absence 
of kaon absorption processes, these channels are uncoupled. This 
makes the three-body approach  closer to the potential multiple 
scattering theory.  However, this statement would be more 
accurate if the MEC contributions would be small. In our paper 
we will show that, indeed, this is case, providing a 
justification of the conventional calculations done so far.

The structure of the paper is the following. In section 2 we 
consider the formalism for the description of the  MEC. Section 
3 presents our results and our conclusions are summarized in 
section 4.   

\section{Formalism}

The conventional approach in the description of the 
$K^+$-nuclear interaction is the potential multiple scattering 
theory. Due to the absence of the absorption channels such 
approach is rather precise in the treatment of the $K^+$-nucleus 
dynamics. In the present paper we follow closely the formalism 
and steps of Ref.\cite{Paez} where $K^+$ scattering on $He$ 
targets was first studied using $S, P, D, F$ partial amplitudes 
for $K^+N$ scattering from Ref.\cite{Martin} and including the 
spin-isospin dependence of the first-order optical potential.  
We calculate this potential in a similar way as it was done for 
pion scattering on the lightest nuclei cite{KTB93,KTB95}. 
Note only that for the deuteron we use realistic wave functions 
obtained with the Paris $NN$ potential~\cite{Lacombe}. In the 
description of $^3He$ ground state we use the solution of the 
Faddeev equation with the Reid-Soft-Core  potential~\cite{Brand}.  
In both cases $S$- and $D$-components of the nuclear wave 
function are taken into account.  $^4He$ is described using a
charge form factor extracted from electron scattering.  The 
contributions from  kaon rescattering are taken into account 
using the KMT version of multiple scattering theory~\cite{KMT}.  
Within this framework the Lippmann-Schwinger integral equation 
is solved with a separable form for the off-shell extrapolation 
of the $KN$ amplitude.  

The main aim of our paper is the study of contributions from MEC 
which are not included in the potential multiple scattering 
theory.  Originally it was supposed that the MEC are caused by 
the $K^+$ interaction with the nuclear pion cloud, i.e. $K^+$ 
scattering from virtual pions exchanged between two-nucleon (see 
Fig. 1a).  The detailed investigations of this mechanism in 
$K^+-^{12}C$ scattering have been done by M. J. Jiang and D. S. 
Koltun \cite{Koltun}. This work was improved in Ref. 
\cite{Oset1} by the addition of extra terms contributing to the 
imaginary part of the $K^+$ selfenergy from the $K^+$ 
interaction with the pion virtual cloud, $\delta \Pi_K$. In 
addition, the static approximation  used to deal with  the 
virtual pion cloud was removed in Ref. \cite{Oset1}, resulting in 
appreciable numerical changes.  Uncertainties remained in the 
real part in \cite{Koltun,Oset1} tied to the off-shell 
extrapolation of the $K\pi$ amplitude. 

However, it was also mentioned in \cite{Koltun} that the 
addition of contact terms from chiral Lagrangians (Fig. 1b) 
should partly cancel the contribution of the real part of 
$\delta\Pi_K$, much as it happens in the $\pi N\rightarrow 
\pi\pi N$ reaction \cite{Oset3,Bernard}, or in the evaluation of 
the pion selfenergy from the interaction of the pion with the 
virtual pion cloud \cite{Roc,Oset4}. A detailed calculation of 
the real part of $\delta \Pi_k$ including the pion pole (Fig. 
1a) and contact term (Fig. 1b) was carried out in Ref. 
\cite{Meiss}  using standard Chiral Lagrangians for the octet of 
pseudoscalar mesons. An exact cancellation was found  for 
symmetric  nuclear matter, unlike in the case of the pion 
selfenergy, where the terms cancelled only partially 
\cite{Roc,Oset4}.

In view of the results obtained in Ref. \cite{Meiss} for the 
real part of $\delta \Pi_K$ and the terms involved in the 
evaluation of the imaginary part of $\delta \Pi_K$, one 
envisages some reduction of the results obtained in Ref. 
\cite{Oset1}, which were already rather small. 

However, the cancellation found in Ref. \cite{Meiss} in nuclear 
matter was for $T=0$ systems  and only because  kaons propagate  
forward in nuclear matter. Away from the forward direction, or if 
$T\neq 0$, this does not occur. It is thus worth  investigating 
in detail whether the effects  of these MEC can show up at large 
angles, as it is usually the case in electromagnetic  MEC.  

In the present work we shall construct the MEC operator starting 
from the $K^+N\rightarrow KN\pi$ amplitude obtained in Ref. 
\cite{Oset2}.  The procedure is similar to the way the  MEC 
operator for photon absorption processes is generated starting 
from the $\gamma N\rightarrow N\pi$ amplitude \cite{Oset5}, or 
for pion photoproduction starting from the $\gamma N\rightarrow 
N\pi\pi$ amplitude \cite{Gomez}. First, using results of 
Ref.\cite{Oset2} we define the effective $NNKK\pi$ vertex 
$t_{eff.}$:  
\begin{eqnarray} t_{eff.}\,=\,{\textstyle 
\frac{1}{2}}\,t_{pole}\,+\,t_{cont.}\,, 
\end{eqnarray} 
where $t_{pole}$ and $t_{cont}$ are  the pion pole and contact terms.  
In Fig. 1 they correspond to Fig. 1a and Fig.1b, respectively, 
without the nucleon line to the right. In a second step we 
assume that the pion produced is off-shell and we let it be 
absorbed by a second nucleon.  Finally, taking into account all 
the possible isospin channels we obtain MEC operators in four 
effective vertices $(t_1,...,t_4)$.  They are associated with 
the diagrams depicted in Fig. 2.  In the evaluation of the 
spin-isospin matrix elements from the two-body MEC operator we 
shall use the method which was suggested in Ref.\cite{Gomez}. It 
consists in writing explicitly the wave function in terms of 
spin and isospin up and down  states and evaluating the matrix 
elements of the operators in that base. Then, neglecting 
the nuclear $D$-states (we expect that due to momentum 
sharing arguments their contributions are small and in any case 
they are corrections in corrections) we get for different 
initial and final nuclear spin states:\\ 
for the $K^+d$ scattering 
\begin{mathletters}
\begin{eqnarray}
A \equiv 2<d+ \mid t_{eff.}(1)\,\vec{\sigma}_2\cdot\vec{p}\,
\tau^{\lambda}_2\mid d+> = -<\uparrow\mid t_d \mid \uparrow > 
< \uparrow \mid \vec{\sigma} \cdot \vec{p} \mid \uparrow > \,,
\end{eqnarray}
\begin{eqnarray}
B & \equiv & 2<d0 \mid t_{eff.}(1)\,\vec{\sigma}_2\cdot\vec{p}\,
\tau^{\lambda}_2\mid d+>= \nonumber\\ & &
-\frac{1}{\sqrt{2}}\left(<\uparrow\mid t_d \mid \uparrow > 
< \downarrow \mid \vec{\sigma} \cdot \vec{p} \mid \uparrow > 
+<\downarrow\mid t_d \mid \uparrow > 
<\uparrow \mid \vec{\sigma} \cdot \vec{p} \mid \uparrow > \right)\,,
\end{eqnarray}
\begin{eqnarray}
C \equiv 2<d- \mid t_{eff.}(1)\,\vec{\sigma}_2\cdot\vec{p}\,\tau^{\lambda}_2\,
\mid d+>=-<\downarrow\mid t_d \mid \uparrow > 
< \downarrow \mid \vec{\sigma} \cdot \vec{p} \mid \uparrow > \,,
\end{eqnarray}
\begin{eqnarray}
D & \equiv & 2<d+ \mid t_{eff.}(1)\,\vec{\sigma}_2\cdot\vec{p}\,
\tau^{\lambda}_2\,\mid d0>= \nonumber\\ & & 
-\frac{1}{\sqrt{2}}\left(<\uparrow\mid t_d \mid \uparrow > 
< \uparrow \mid \vec{\sigma} \cdot \vec{p} \mid \downarrow > 
+<\uparrow\mid t_d \mid \downarrow > 
<\uparrow \mid \vec{\sigma} \cdot \vec{p} \mid \uparrow > \right)\,,
\end{eqnarray}
\begin{eqnarray}
E & \equiv & 2<d0 \mid t_{eff.}(1)\,\vec{\sigma}_2\cdot\vec{p}\,
\tau^{\lambda}_2\,\mid d0>= \nonumber\\ & & 
-A-\frac{1}{2}\left(<\uparrow\mid t_d \mid \downarrow > 
< \downarrow \mid \vec{\sigma} \cdot \vec{p} \mid \uparrow > 
+<\downarrow\mid t_d \mid \uparrow > 
<\uparrow \mid \vec{\sigma} \cdot \vec{p} \mid \downarrow > \right)\,,
\end{eqnarray}
\end{mathletters}
for the $K^+\,^3He$ scattering 
\begin{mathletters}
\begin{eqnarray}
F & \equiv &  6<^3He\uparrow \mid t_{eff.}(1)\,\vec{\sigma}_2\cdot\vec{p}\,
\tau^{\lambda}_2 \mid ^3He\uparrow>= 
-<\uparrow\mid 2t_1+ t_{24}^{(+)} \mid \uparrow > 
< \uparrow \mid \vec{\sigma} \cdot \vec{p} \mid \uparrow > 
- \nonumber\\ & & -
<\uparrow\mid t_1+\sqrt{2}t_2 \mid \downarrow > < \downarrow \mid 
\vec{\sigma} \cdot \vec{p} \mid \uparrow > 
-<\downarrow\mid t_1+\sqrt{2}t_4 \mid \uparrow > < \uparrow \mid 
\vec{\sigma} \cdot \vec{p} \mid \downarrow > \,,
\end{eqnarray}
\begin{eqnarray}
G & \equiv & 6<^3He\downarrow \mid t_{eff.}(1)\,\vec{\sigma}_2\cdot\vec{p}\,
\tau^{\lambda}_2 \mid ^3He\uparrow>= \nonumber\\ & &  
-<\downarrow\mid t_{24}^{(-)} \mid \uparrow > < \uparrow \mid \vec{\sigma} 
\cdot \vec{p} \mid \uparrow > -<\downarrow\mid t_{24}^{(-)} \mid \downarrow > 
< \downarrow \mid \vec{\sigma} \cdot \vec{p} \mid \uparrow >\,. 
\end{eqnarray}
\end{mathletters}
In Eqs. (2-3) $\mid\uparrow>$ and $\mid\downarrow>$ 
are the up and down  spin states of the single 
nucleon.  The operators $t_d$ and $t_{24}^{(\pm)}$ are the 
following combinations of the effective vertices 
\begin{equation} 
t_d\,=\,t_1\,-\,t_3\,+\,t_{24}^{(+)}\,,\quad\quad 
t_{24}^{(\pm)}=\sqrt{2}(t_2\pm t_4)\,.  
\end{equation} 
Some details for the calculation of the matrix elements (2-3) are 
given in the Appendix. 

One should note that the diagrams in Fig. 1a and Fig. 1b involve 
a model for the $K^+ N\rightarrow K^+N\pi$ process based on the 
pion pole and contact terms. The model of Ref. \cite{Oset2} 
contains also a term for $\Delta$ excitation which was proved 
important in the $K^+ N\rightarrow K N\pi$ reaction in Ref. 
\cite{Bland}. This term would lead to the MEC term of Fig. 3. 
However, in the case of the deuteron this term is zero because 
the intermediate $\Delta N$ state cannot have isospin zero like 
the deuteron. In the case of  $^3He$ and $^4He$, and considering 
only $S$-wave nuclear states, we also find that the $\Delta$ 
excitation term does not contribute  to the non-spin-flip 
amplitude. However, it contributes to the spin-flip amplitude 
$G$ in the case of  $^3He$, but we find it negligible compared 
to the MEC related to the chiral terms.

The final expression for the MEC amplitude, which has to be added 
to the first-order kaon-nuclear potential, is given  by
\begin{equation}
V_{MEC}^{fi}=-\frac{E_A(\vec{k})}{4\pi E}\,\int\frac{d\vec{p}}{(2\pi)^3}\,
\frac{F_A(\vec{Q},\vec{p})}{\vec{p}\,^2+m_{\pi}^2}
T_{fi}^{(2)}(\vec{k},\vec{k}\,',\vec{p})\,F(\vec{p})\,F(\vec{Q}-\vec{p})
\frac{f_{\pi}}{m_{\pi}}\,,
\end{equation}
where $f_{\pi}^2/4\pi=0.08$ is the $\pi NN$ coupling constant, 
$T_{fi}^{(2)}$ are the spin-isospin matrix elements $A,...,G\,$ 
from Eqs. (2) and (3), $E_A(\vec{k})$ and 
$E=E_K(\vec{k})+E_A(\vec{k})$ are the kaon 
and total kaon-nuclear energies in the c.m. frame, $F(\vec{q})$ 
is the off-shell form factor: 
$F(\vec{q})=(\Lambda^2-m_{\pi}^2)/(\Lambda^2+\vec{q}\,^2)$ 
with $\Lambda$=1300 MeV, $\vec{Q}=\vec{k}-\vec{k}\,'$ is the transferred
momentum.
 
The two-body nuclear form factor $F_A(\vec{Q},\vec{p})$ in the case
of the deuteron is given by
\begin{equation}
F_d(\vec{Q},\vec{p})=\int u_S^2(x)\,j_0(\mid \vec{p}-
{\textstyle \frac{1}{2}}\vec{Q}\mid x)\,dx\,,
\end{equation}
where $u_S(x)$ is the $S$-wave part of the deuteron wave function 
taken from Ref.\cite{Lacombe}. For the $^3He$ we have
\begin{equation}
F_{^3He}(\vec{Q},\vec{p})=\int u_S^2(x,y)\,j_0(\mid \vec{p}-
{\textstyle \frac{1}{2}}\vec{Q}\mid x)\,j_0({\textstyle 
\frac{1}{3}}Qy)\,dx\,dy\,,
\end{equation}
where $x$ and $y$ are the standard Jacobi coordinates. 
$u_S(x,y)$ is the $S$-wave part of the radial wave function for 
the trinucleon system obtained in Ref.\cite{Brand} by solving the 
Faddeev equation.  In the case of a simple harmonic 
oscillator model the two-body form factor can be expressed in 
the analytical form 
$F_{^3He}(\vec{Q},\vec{p})=e^{-b^2Q^2/6}\,e^{-b^2\vec{p}\cdot 
(\vec{p}+ \vec{Q})/2}$.

On finishing this section let us make a few comments about the MEC. 
First, it turns out that in the case of $K^+\,d$ scattering in 
the forward direction the MEC contributions from the contact and 
pion pole terms cancel each other. This result is consistent 
with the results of Ref.\cite{Meiss} where this effect was 
proved for a more general case:  $K^+$ scattering with symmetric 
nuclear matter.  The next comment is connected with the 
spin-flip transition in kaon scattering on $^3He$.  The 
one-body part  of the spin-flip amplitude is proportional to 
$\sin\theta_K$. The same angle dependence was also obtained for 
the MEC spin-flip amplitude. Note that this amplitude contains 
only the $t_2$ and $t_4$ effective vertices. This mean that there 
are no contributions from $\pi^0$-exchange diagrams.

The last comment concerns the method used here for  deriving  
the nuclear matrix elements of the two-body operator. It is 
similar to the method used in quark models. On the other hand 
there is an alternative way based on the algebra of tensor 
operators. For the MEC we have done calculations with both 
methods in order to have extra confidence in the results, which 
were identical in both cases.

\section{Results and Discussion}

We begin with the consideration of $K^+d$ scattering.  In 
the analysis of the experimental data the simple expressions obtained 
in Plane Wave Impulse Approximation (PWIA), with only deuteron 
$S$-state, are normally used in order to extract information about 
kaon-neutron scattering amplitude. However, this is 
an approximate expression where the deuteron $D$-state and 
rescattering corrections are neglected. In Fig. 4 we 
demonstrate the accuracy of such an approximation. In fact, it 
is good at momentum transfers $Q^2=-t<0.2$ (GeV/c)$^2$ and it 
turns out that most of the analysed experimental points are 
concentrated in this region.  This means that the results which 
could be obtained using more accurate expressions would lead to 
small changes in the kaon-neutron amplitude. We come to the same 
conclusion in the analysis of the total cross section.

Another correction which is not taken into account in the simple 
PWIA approach is the effect from MEC coming from diagrams  
depicted in Figs. 1 and 2. We have found (see Fig. 5 ) that in 
$K^+-d$ elastic scattering their contributions are small in 
comparison with contribution from the conventional one-body 
mechanism. For example, due to the cancelation effects found in 
Ref.\cite{Meiss}, there is no MEC contribution in the forward 
direction.  At backward angles due to the interference with the 
one-body (PWIA) amplitudes  they reduce the differential cross 
section in about a factor of two, but this effect is practically 
cancelled by contributions from kaon rescattering.  The results 
for the $K^+-^4He$ scattering are similar.  

In $K^+-^3He$ scattering (see Fig. 6), in contrast to the 
deuteron and $^4He$ cases, the contribution from MEC in the forward 
direction is not zero.  However, in this region the 
conventional one-body mechanism dominates.  At backward angles 
the contribution from kaon rescattering is more important. 
We have also found that in this region it is very important to 
use realistic Faddeev wave functions instead of the simple 
harmonic oscillator model.  In the case of MEC this essentially 
enhances the contribution of the non-spin-flip amplitude.

The MEC mechanisms which we are using here correspond to the 
consideration of the real part of the $K^+$ selfenergy from the 
interaction of the kaon with the virtual  pion cloud. The 
corrections from the imaginary part  are very small at low 
energies and were found to provide corrections of 10-20\% at 
$K^+$ momenta around 800 MeV/c  for $^{12}C$. In the lightest 
nuclei considered here  the effects should be certainly smaller. 
Also, as indicated above, the consideration of the findings of 
Ref. \cite{Meiss} would lead to further reductions in the 
imaginary part  of $\delta \Pi$, and we should expect the 
corrections from the imaginary part to be small, as the one 
found here from the real part, essentially negligible within the 
accuracy of the present data and the possible one in a near 
future. On the other hand, the MEC evaluated here would account 
for the $2p2h$ excitation diagrams considered in Ref. 
\cite{Meiss}. In addition one would have contributions from $ph 
\Delta h$ components which are of the same order of magnitude as 
those of the $2p2h$ (and of opposite sign). In view  of the 
smallness of the contributions obtained from the $2p2h$ terms  
we refrain from extending the calculations to account for these 
corrections  which would not change the conclusion drawn  here, 
i.e.  the little relevance of the MEC in this problem compared 
to the one-body and rescattering contributions.

\section{Conclusion}

The $K^+$ scattering on the deuteron is the main source of 
information about the $K^+$-neutron  interaction. The basic 
theoretical approach which is normally used for this purpose is 
the simple Plane Wave Impulse Approximation  which does not 
take into account contributions from the D-state components of 
the nuclear wave functions, from kaon rescattering and 
Meson Exchange Currents. In the present paper we have analysed 
these ingredients including all of them together.  At the 
same time we have evaluated cross sections for $^3He$ and 
$^4He$ targets using these ingredients. 

The MEC operator for the finite nuclei was constructed for the 
first time using as an input the amplitude for the 
$K^+N\rightarrow K^+N\pi$ reaction, which was obtained in 
Ref.\cite{Oset2} in the framework of the Standard Chiral 
Perturbation Theory. We have found that MEC contributions are 
small at momentum transfers $Q^2<0.5$ (Gev/c)$^2$. In this 
region more important corrections are coming from kaon 
rescattering, especially in the total cross section at low 
energies ($k_{lab}<$ 500 MeV). From this result we can conclude 
that the conventional multiple scattering theory which does not 
include MEC contributions is a reliable approach for kaon 
scattering on the lightest nuclei in this kinematical domain.

At larger momentum transfers the MEC can become more relevant 
and, for instance, for $K^+$ scattering in the deuteron and 
$^3He$, with kaon momentum around 600 MeV/c  and backward 
angles, they can reduce the cross section in about a factor of 
two. However, the cross sections in that region are about four and five 
orders of magnitude, respectively, smaller than at forward 
angles.  In the case of $K^+$ scattering on the deuteron we 
have demonstrated that at $k_{lab}<$400 MeV/c and momentum 
transfers $Q^2<$0.2 (GeV/c)$^2$, where most of the analysed 
experimental data are concentrated, the corrections from the 
deuteron $D$-state and kaon rescattering are small.  Therefore, 
the kaon-neutron scattering amplitude that one would obtain 
using our more accurate approach would be the same one obtained 
so far.  

\acknowledgments
This work is partially supported by the CICYT contract no. 
AEN-96-1719.  One of us (S.S.K.) wants to acknowledge support 
from the Ministerio de Educacion y Ciencia in his sabbatical 
stay at the University of Valencia and to say many thanks to 
Juan el Rey whose help allowed us to finish this work.

\appendix
\section*{}

In this Appendix we present some details for the calculations of 
the MEC.  First, let us consider expressions for the four 
effective vertices $t_1,...,t_4$ which correspond to the diagrams 
depicted in Fig. 2 without the nucleon line to the right. The 
vertices $t_1$ and $t_2$, which describe the $(K^+,K\pi)$ 
reaction on the proton, were considered in detail in 
Ref.\cite{Oset2}. The other vertices $t_3$ and $t_4$ are new. They 
describe the $(K^+,K\pi)$ reaction on the neutron. As we 
discussed in  Section 2, all these vertices involve the 
contribution from the pion pole and contact terms. The explicit 
expressions for them are the following (in the notation of 
Ref.\cite{Oset2}):\\ 
Pion pole terms  
\begin{eqnarray}
t_1^p=2\,C_1\frac{<p\mid \gamma^{\mu}\gamma_5\,\tau_3\mid p>}
{q^2-m_{\pi}}\,q_{\mu}\,\{2m_K^2-2k\cdot k\,'+p\cdot q+m_{\pi}^2\}\,,
\end{eqnarray}
\begin{eqnarray}
t_2^p=4\sqrt{2}\,C_1\frac{<n\mid \gamma^{\mu}\gamma_5\,\tau_-\mid p>}  
{q^2-m_{\pi}}\,q_{\mu}\,\{m_K^2-p\cdot k-2p\cdot k\,'-k\cdot k\,'
+m_{\pi}^2-p^2\}\,,
\end{eqnarray}
\begin{eqnarray}
t_3^p=2\,C_1 \frac{<n\mid \gamma^{\mu}\gamma_5\,\tau_3\mid n>}
{q^2-m_{\pi}}\,q_{\mu}\,\{2m_K^2-2k\cdot k\,'+p\cdot q+m_{\pi}^2\}\,,
\end{eqnarray}
\begin{eqnarray}
t_4^p=4\sqrt{2}\,C_1 \frac{<p\mid \gamma^{\mu}\gamma_5\,\tau_+\mid n>}
{q^2-m_{\pi}}\,q_{\mu}\, \{m_K^2-q\cdot k-2q\cdot k\,'-k\cdot k\,'
+m_{\pi}^2-q^2\}\,.
\end{eqnarray}
Contact terms
\begin{eqnarray}
t_1^c=(C_1-C_2)<p\mid \gamma^{\mu}\gamma_5\,\tau_3\mid p>  
\{k_{\mu}'-k_{\mu}-2p_{\mu}\}\,,
\end{eqnarray}
\begin{eqnarray}
t_2^c=-\sqrt{2}\,C_1<n\mid \gamma^{\mu}\gamma_5\,\tau_-\mid p>  
\{p_{\mu}+k_{\mu}'+2k_{\mu}\}\,,
\end{eqnarray}
\begin{eqnarray}
t_3^c=C_2<n\mid \gamma^{\mu}\gamma_5\,\tau_3\mid n>  
\{k_{\mu}'-k_{\mu}-2p_{\mu}\}\,,
\end{eqnarray}
\begin{eqnarray}
t_4^c=-\sqrt{2}\,C_1<p\mid \gamma^{\mu}\gamma_5\,\tau_+\mid n>  
\{p_{\mu}-2k_{\mu}'-k_{\mu}\}\,,
\end{eqnarray}
where $\mid p>$ and $\mid n>$ are the proton and neutron states, 
respectively, $\tau_+$ and $\tau_-$ are the standard raising and 
lowering isospin operators ($\tau_+\mid n>=\mid p>$ and 
$\tau_-\mid p>=\mid n>$).  In Eqs. (A1-A8) the coupling 
constants $C_1$ and $C_2$ are defined as
\begin{eqnarray}
C_1=\frac{D+F}{2}\frac{1}{12 f^3},\qquad\qquad
C_2=\frac{D-F}{2}\frac{1}{12 f^3}
\end{eqnarray}
with $D=0.85,\, F=0.52$ and $f=92.4$MeV. Recall also the 
convention used for the momentum conservation for the pion pole 
term:  $\vec{k}=\vec{k}\,'+\vec{p}+\vec{q}$. 

 In order to construct the MEC operator in the nuclear application 
we use a nonrelativistic expression for the $\pi NN$ 
vertex, i.e.  $<N\mid \gamma^{\mu}\gamma_5\mid 
N>q_{\mu}=-\vec{\sigma}\cdot\vec{q}$ and take into account that  
pions are off-shell ($q_0=p_0=0$).  Finally after some algebra 
we obtain the following expression for the operator $t_d$,  
defined in Eq. (4), and which describes the MEC in kaon deuteron 
elastic scattering:
\begin{eqnarray}
t_d= 3\,C_1\,\left[2\,\frac{m_{\pi}^2-\vec{p}\cdot \vec{q}-\vec{Q}^2}
{\vec{q}\,^2+m_{\pi}^2}\,\vec{\sigma}\cdot\vec{q}\,+\,
\vec{\sigma}\cdot(2\vec{p}+\vec{Q})\right]\,.
\end{eqnarray}
Using this simple expression the matrix elements $A,...,E$ from 
Eqs. (2) can be easily derived. From Eq. (A10) we can also see 
that in the forward direction, where the momentum transfer is 
$\vec{Q}=\vec{k}- \vec{k}\,'=0$ and $\vec{p}=-\vec{q}$, the pion 
pole term is cancelled by the contact term.

In the case of kaon scattering on $^3He$  we have another 
combination of $t_1,...,t_4$ vertices (see Eqs. (3)). They can 
be derived in a similar way. Here we present only the final 
result for the non-spin-flip $F$ and spin-flip $G$ matrix 
elements from Eq. (3):   
\begin{eqnarray}
F=-6\,C_1\,\frac{m_{\pi}^2-\vec{p}\cdot \vec{q}-\vec{Q}^2}
{\vec{q}\,^2+m_{\pi}^2}\,\vec{p}\cdot\vec{q} -
2(2C_1-C_2)(2\vec{p}\,^2+\vec{Q}\cdot\vec{p})\,,
\end{eqnarray}
\begin{eqnarray}
G=6\sqrt{2}\,C_1\left\{2\,\,\frac{\vec{p}\cdot (\vec{k}+\vec{k}\,')}
{\vec{q}\,^2+m_{\pi}^2}\,i[\vec{p}\times\vec{q}\,]_{+1} - 
i[\vec{p}\times(\vec{k}+\vec{k}\,')]_{+1}\right\}\,,
\end{eqnarray}
where the product $i[\vec{A}\times\vec{B}]_{+1}=A_{+1}B_0- 
A_{0}B_{+1}$ is defined in the covariant spherical basis.  Note 
that expressions (A10-A11) for the contributions of the pion 
terms, together with Eq. (5), are symmetrical relative to the  
permutation $\vec{p} \leftrightarrow \vec{q}$.

%%%%%%%%%%%   Fig. 1 %%%%%%%%%%%%
\begin{figure}   
\caption{ A diagram for MEC in kaon-nucleus scattering: {\bf (a)} pion
pole term, {\bf (b)} contact term.}
\end{figure}

%%%%%%%%%%%   Fig. 2 %%%%%%%%%%%%
\begin{figure}
\caption{ Diagrammatic representation of the MEC for the four 
isospin channels in terms of the effective $NNKK\pi$ vertices 
$t_1,...,t_4$. In diagram (a) the proton line $p$ to the right 
contributes only in the $^3He$ case.}
\end{figure}

%%%%%%%%%%%   Fig. 3 %%%%%%%%%%%%
\begin{figure}
\caption{  $\Delta$ excitation term in the MEC}
\end{figure}

%%%%%%%%%%%   Fig. 4 %%%%%%%%%%%%
\begin{figure}
\caption{ Elastic differential cross sections $d\sigma/dt$ for 
$K^+d$ scattering  at lab. kaon momenta $k_{lab.}=$ 342, 587 and 
890 MeV/c.  The dotted and dashed curves are the PWIA 
calculations without and with deuteron $D$-state, respectively. 
The solid curves are the full calculations (with kaon 
rescattering and Coulomb interaction). Experimental data are 
from Ref.\protect\cite{Sakitt} (o) and 
Ref.\protect\cite{Glasser} $(\bullet)$.}
\end{figure}

%%%%%%%%%%%   Fig. 5 %%%%%%%%%%%%
\begin{figure}
\caption{ MEC effects in elastic $K^+d$ scattering at 
$k_{lab.}=$ 587 MeV/c.  The dashed and dash-dotted curves are 
PWIA and PWIA + MEC results.  The solid curve is the result of 
the full calculations.  By the dotted curve we denote the MEC 
contribution alone. Experimental data are from 
Ref.\protect\cite{Sakitt} (o) and Ref.\protect\cite{Glasser} 
$(\bullet)$.}
\end{figure}

%%%%%%%%%%%   Fig. 6 %%%%%%%%%%%%
\begin{figure}
\caption{ The same as in Fig. 5, but for elastic $K^+-^3He$ 
scattering calculated using Faddeev wave function from 
Ref.\protect\cite{Brand}.} 
\end{figure}


\begin{references}
\bibitem[*]{Sabit}
Permanent address: Laboratory of Theoretical Physics, JINR Dubna,
Head Post Office Box 79, SU-101000 Moscow, Russia.
\bibitem{SKG} P. B. Siegel, W. B. Kaufmann, and W. R. Gibbs, 
  Phys. Rev. C {\bf 30}, 1256 (1984); C {\bf 31}, 2184 (1985)
\bibitem{Weise} W. Weise, Nuovo Cimento, {\bf 102 A}, 265 (1989)
\bibitem{Chen} C. M. Chen and D. J. Ernst, Phys. Rev. C {\bf 45}, 2011 (1992)
\bibitem{Ernst} D. J. Ernst, M. F. Jiang, C. M. Chen, and Mikkel B. Johnson
{\it Proceedings of the 6th International Conference "Mesons and Light 
Nuclei 95"} Straz pod Ralskem, Cheh Republic, July 3-7, 1995.
\bibitem{Koltun} M. F. Jiang and Daniel S. Koltun, Phys. Rev. C {\bf 46},
2462 (1992)
\bibitem{Oset1} C. Garcia-Recio,J. Nieves, and E. Oset, Phys. Rev. C
{\bf 51}, 237 (1995)
\bibitem{Meiss} Ulf-G. Meissner, E. Oset, and A. Pich, Phys. Lett. 
{\bf B353}, 161 (1995)
\bibitem{Mich} R. Michael {\it et al.}, Phys. Lett. {\bf B382}, 29 (1996) 
\bibitem{Landau75} R. H. Landau, Ann. Phys. (N.Y.) {\bf 92}, 205 (1975)
\bibitem{Mach76} R. Mach, Nucl. Phys. {\bf A256}, 513 (1976)
\bibitem{LT78} R. H. Landau and A. W. Thomas, Nucl. Phys. {\bf A302},
461 (1978)   
\bibitem{KTB93}   S. S. Kamalov, L. Tiator and C. Bennhold,
  Phys. Rev. C {\bf 47}, 941 (1993)
\bibitem{KTB95}   S. S. Kamalov, L. Tiator and C. Bennhold, Preprint 
Mainz, MKPH-T-95-21; Phys. Rev. C (in print)
\bibitem{Paez}   M. J. Paez and R. H. Landau, Phys. Rev. C {\bf 24},
1120 (1981)
\bibitem{Oset2} E. Oset and M. J. Vicente Vacas, Phys. Lett. 
{\bf B386}, 39 (1996)

\bibitem{Martin} B. R. Martin, Nucl. Phys. {\bf B94}, 413 (1975)
\bibitem{Lacombe} M. Lacombe, B. Loisean, R. VinhMau, J. Cote, P. Pires,
and R. de Tourreil, Phys. Lett. B {\bf 101}, 139 (1981)
\bibitem{Brand} R. A. Brandenburg, Y. E. Kim, and A. Tubis, Phys. Rev.
Phys. Rev. C {\bf 12}, 1368 (1975) 
\bibitem{KMT} A. K. Kerman, H. McManus, and R. M. Thaler,
Ann. Phys. (N.Y.) {\bf 8}, 551 (1959)
\bibitem{Oset3} E. Oset and M. J. Vicente-Vacas, Nucl. Phys. {\bf A446},
584 (1985)
\bibitem{Bernard} V. Bernard, N.-Kaiser and Ulf-G. Meissner,
Nucl. Phys. {\bf B457}, 147 (1995)  
\bibitem{Roc} R. Rockmore, Phys. Rev. C {\bf 40}, R13 (1989)
\bibitem{Oset4} E. Oset, G. Garcia-Recio and J. Nieves, Nucl. Phys.
{\bf A584}, 653 (1995)
\bibitem{Oset5} R. C. Carrasco and E. Oset, Nucl. Phys. {\bf A536}, 445 (1992) 
\bibitem{Gomez} J. A. Gomez-Tejedor, S. S. Kamalov, and E. Oset, 
Phys. Rev. C {\bf 54}, 3160 (1996) 
\bibitem{Bland} R. W. Bland {\it et al.}, Nucl. Phys. {\bf B13}, 595 (1969);
{\it ibid.} Nucl. Phys. {\bf B18}, 537 (1970)
%\bibitem{Bugg} D. V. Bugg. {\it et al.}, Phys. Rev. {\bf 168}, 1466 (1968)
%\bibitem{Marlow} D. Marlow {\it et al.}, Phys. Rev. C {\bf 25}, 2619 (1982)
\bibitem{Sakitt} M. Sakitt, J. Skelly, and J. A. Thompson, 
 Phys. Rev. D {\bf 12}, 3386 (1975)
\bibitem{Glasser} R. G. Glasser {\it et al.} Phys. Rev. D {\bf 15}, 1200 (1977) 
\end{references}
\end{document}